\documentclass[fleqn,usenatbib]{mnras}
\usepackage{footnote,graphicx,natbib,color,multirow,amsmath,url,amssymb,tabularx,amssymb, mathtools, listings, float, setspace}

\usepackage{newtxtext,newtxmath, color}

\usepackage{aas_macros}
\usepackage{ragged2e}

\usepackage[T1]{fontenc}

\DeclareRobustCommand{\VAN}[3]{#2}
\let\VANthebibliography\thebibliography
\def\thebibliography{\DeclareRobustCommand{\VAN}[3]{##3}\VANthebibliography}

\usepackage{colortbl}
\usepackage[table]{xcolor}
\definecolor{lightgray}{gray}{0.95}

\usepackage{longtable}
\usepackage{array} 

\newcolumntype{L}{>{\raggedright\arraybackslash}}
\newcolumntype{R}{>{\raggedleft\arraybackslash}}
\newcolumntype{C}{>{\centering\arraybackslash}}

\renewcommand{\tabcolsep}{0.2cm}

\def\lesssim{\mathrel{\hbox{\rlap{\hbox{\lower3pt\hbox{$\sim$}}}\hbox{\raise2pt\hbox{$<$}}}}}

\definecolor{check}{rgb}{0,0,0}

\def\oiii		{$\mathrm{\left[ O \textsc{iii}\right] }$}

\def\sii		{$\mathrm{\left[ S \textsc{ii}\right] }$}

\def\ha               {H$\alpha$}

\def\hbeta           {$\rm{H}\beta$}

\definecolor{check}{rgb}{0,0,0}

\definecolor{referee}{rgb}{0,0,0}
\def\refchange		{\color{referee}}

\definecolor{referee2}{rgb}{0,0,0}

\title[Non-merger driven AGN Outflows]{Kiloparsec-scale AGN Outflows and Feedback in Merger-Free Galaxies}
\author[Smethurst et al. 2021]{R.~J.~Smethurst,$^{1}$\thanks{E-mail: rebecca.smethurst@physics.ox.ac.uk} B.~D.~Simmons,$^{2, 3}$ A.~Coil,$^{3}$ C.~J.~Lintott,$^{1}$  W.~Keel,$^{4}$ K. L. Masters,$^{5}$ E. Glikman,$^{6}$ \newauthor G.~C.~K.~Leung,$^{3,7}$  J.~Shanahan,$^{3}$ I. L. Garland$^{2}$
\\ 
$^1$ Oxford Astrophysics, Department of Physics, University of Oxford, Denys Wilkinson Building, Keble Road, Oxford, OX1 3RH, UK \\ 
$^2$ Physics Department, Lancaster University, Lancaster, LA1 4YB, UK \\
$^3$ Center for Astrophysics and Space Sciences (CASS), Department of Physics, University of California, San Diego, CA 92093, USA \\ 
$^4$ Department of Physics and Astronomy, University of Alabama, Box 870324, Tuscaloosa, AL 35487, USA \\ 
$^5$ Departments of Physics and Astronomy, Haverford College, 370 Lancaster Avenue, Haverford, Pennsylvania 19041, USA\\
$^6$ Department of Physics, Middlebury College, Middlebury, VT 05753, USA\\
$^7$ Department of Astronomy, The University of Texas at Austin, Austin, TX 78712, USA\\
}

\date{Accepted 10th August 2021. Received 6th August 2021; in original form 7th July 2021}
\pubyear{2021}

\begin{document}
\label{firstpage}
\pagerange{\pageref{firstpage}--\pageref{lastpage}}
\maketitle

\begin{abstract}

Recent observations and simulations have challenged the long-held paradigm that mergers are the dominant mechanism driving the growth of both galaxies and supermassive black holes (SMBH), in favour of non-merger (secular) processes. In this pilot study of merger-free SMBH and galaxy growth, we use Keck Cosmic Web Imager spectral observations to examine four low-redshift ($0.043 < z < 0.073$) disk-dominated `bulgeless' galaxies hosting luminous AGN, assumed to be merger-free. We detect blueshifted broadened [OIII] emission from outflows in all four sources, which the \oiii/\hbeta~ratios reveal are ionised by the AGN. We calculate outflow rates in the range $0.12-0.7~\rm{M}_{\odot}~\rm{yr}^{-1}$, with velocities of $675-1710~\rm{km}~\rm{s}^{-1}$, large radial extents of $0.6-2.4~\rm{kpc}$, and SMBH accretion rates of $0.02-0.07~\rm{M}_{\odot}~\rm{yr}^{-1}$. We find that the outflow rates, kinematics, and energy injection rates are typical of the wider population of low-redshift AGN, and have velocities exceeding the galaxy escape velocity by a factor of $\sim30$, suggesting that these outflows will have a substantial impact through AGN feedback. Therefore, if both merger-driven and non-merger-driven SMBH growth lead to co-evolution, this suggests that co-evolution is regulated by feedback in both scenarios. Simulations find that bars and spiral arms can drive inflows to galactic centres at rates an order of magnitude larger than the combined SMBH accretion and outflow rates of our four targets. This work therefore provides further evidence that non-merger processes are sufficient to fuel SMBH growth and AGN outflows in disk galaxies.  


\end{abstract}


\section{Introduction}

Determining the physical processes that drive the growth of both galaxies and their supermassive black holes (SMBHs) is a key goal of current observational and theoretical work \citep[see][for a review]{heckmanbest14}. An increasing body of evidence shows that galaxy growth mainly occurs through `secular processes' rather than by mergers. For example, \cite{kaviraj13} show that only $27\%$ of star formation is triggered by major or minor mergers at $z\sim2$, the peak of both star formation and black hole accretion activity.  In addition, \cite{parry09} find that in the Millenium simulations that only $\sim 35\%$ of bulge mass is built by mergers, with the majority built through disk instabilities (triggered through interactions with nearby satellites). Similarly, many recent results have pointed to secular processes as the main driver of SMBH growth. For example, \cite{martin18} showed that in their hydro-dynamical simulations (RAMSES) only $35\%$ of the cumulative growth of SMBHs since $z\sim3$ could be attributed to mergers, both major and minor. Similarly \cite{mcalpine20} found in the EAGLE simulations that galaxy mergers do not induce a significant amount of black hole growth yet do increase the rate of luminous AGN, concluding that that on average no more than $15$\% of a SMBHs mass at $z\sim0$ comes from the enhanced accretion rates triggered by a merger. 

These results, among others both observational and theoretical, challenge the long accepted paradigm whereby mergers are responsible for the correlations between SMBHs and bulges, such as velocity dispersion and bulge mass \citep{magorrian98, haringrix04, vdb16, batiste17, davis19}. Galaxies which have evolved via mergers are easily recognisable, as mergers are able to redistribute angular momentum in galaxy systems, transferring stars from rotation-supported orbits to pressure-supported orbits in a central bulge, similar to an elliptical galaxy. While there is also an increasing number of simulations finding that a disk can reform post-gas rich merger \citep{hopkins09c, sparre17, pontzen17,peschken20, jackson20}, a significant bulge component still forms even in a minor merger \citep[i.e. when the mass ratio in the merger exceeds $10:1$;][]{walker96, hopkins12, tonini16, stevens16}. Therefore, galaxies with little, to no bulge, can be assumed to have merger-free (and interaction-free) histories, at least since $z\sim2$ \citep{martig12}. The growth of both the galaxy and the SMBH in such systems, will have been dominated by non-merger processes alone. 

\citet*[hereafter SSL17]{ssl17} calculated the masses of SMBHs powering such a sample of $101$ disk-dominated AGN and showed that they were over-massive (up to $\sim2$ dex) than would be expected from the black hole-bulge mass relation of \citet{haringrix04}. However, SSL17 also found that their disk-dominated AGN still lay on the total stellar mass-SMBH mass relation. This result suggested that secular processes were able to grow a SMBH at rates higher than previously thought. 

\citet[hereafter S19]{smethurst19b} investigated these possible growth rates by measuring the \oiii~outflow rates in $12$ disk-dominated galaxies using narrowband imaging from the Shane-3m telescope at the Lick Observatory. Under the assumption that the inflow rate to the AGN will be at least equal to the sum of the the outflow rate and the SMBH accretion rate, S19 found that the inflow rates they inferred could be achieved by non-merger processes, including funnelling of gas by bars and spiral arms, and cold accretion from the surrounding galaxy halo. However, this work was limited by the inability to adequately distinguish between gas ionised by the AGN outflow and star formation within the galaxy, and the subtraction of the central AGN PSF (leading to an overestimate and underestimate of the outflow rate respectively).   

In this work, we aim to measure the outflow rates in 4 of the galaxies observed in S19 using spectral observations taken with the Keck Cosmic Web Imager (KCWI). High spectral resolution observations allow for the narrow component in \oiii~(ionised by star formation or the central AGN) to be isolated from the broad component in \oiii~(assumed to be ionised by the AGN outflow). This allows us to derive the outflow rate in these systems more accurately than in the previous study of S19. By using a sample of galaxies where we can be sure that secular processes dominate, we can isolate the merger-free growth path and understand the limitations to merger-free SMBH growth.

In the rest of this work we adopt the Planck 2015 \citep{planck16} cosmological parameters with $(\Omega_m, \Omega_{\lambda}, h) = (0.31, 0.69, 0.68)$ and any emission or absorption features referred to are in the Lick system. All uncertainties on calculated values are determined in quadrature, and all uncertainties on quoted mean values are the standard error on the mean.  In Section~\ref{sec:sample} discuss our sample selection and in Section~\ref{sec:obs} describe our observations. In Section~\ref{sec:data} we describe our data reduction and analysis process, including how we determine the outflow rates in these systems. In Section~\ref{sec:results} we state our results and discuss their implications in Section~\ref{sec:discussion}. Finally, we summarise our conclusions in Section~\ref{sec:conc}.

\begin{figure*}
\centering
\includegraphics[width=\textwidth]{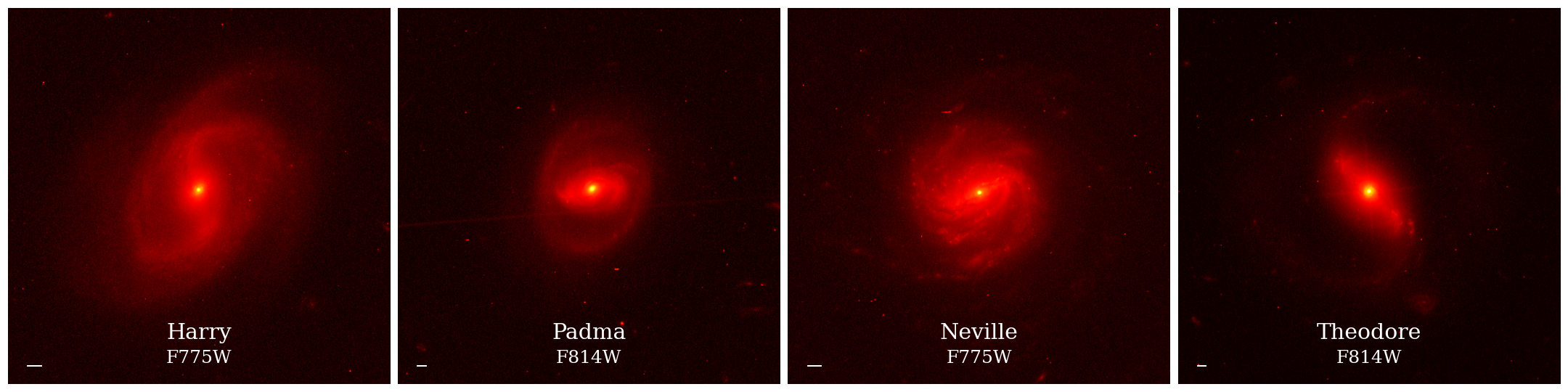}
\caption{\emph{HST} ACS WFC postage stamp images of the $4$ disk-dominated AGN observed with KCWI. North is up and a stretch of 0.55 ($Q=12$) is applied. In each image the HST filter is noted. The AGN can be seen as a bright point source in the centre of each image, which we assume is powered by merger-free processes due to the disk-dominated morphology of these sources. The white bars show $1~\rm{kpc}$ for scale in each panel.} 
\label{fig:hsttargets}
\end{figure*}

\begin{figure*}
\centering
\includegraphics[width=\textwidth]{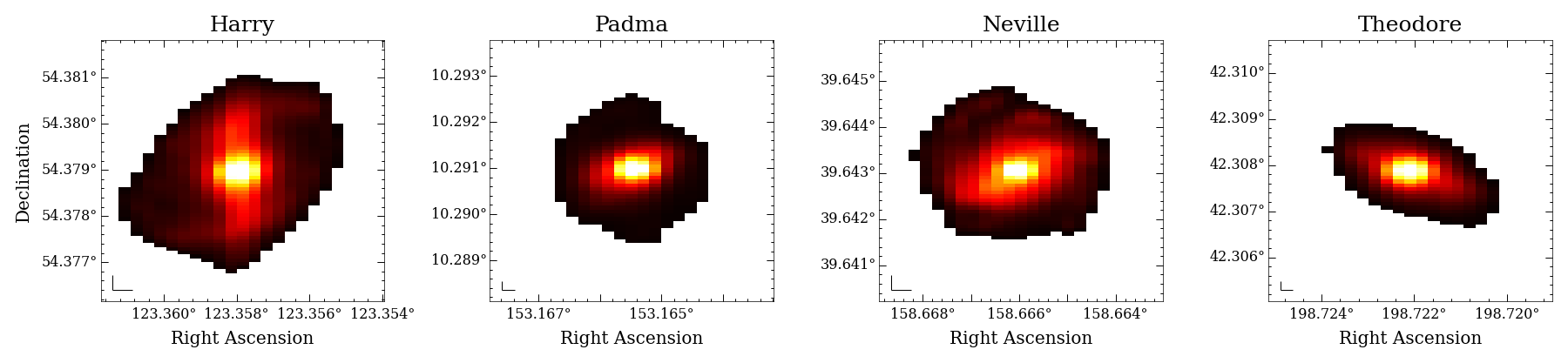}
\caption{Total integrated flux across the IFU data cubes of the $4$ disk dominated AGN observed with KCWI. North is up and an arcsinh stretch is applied. The bar features seen in the HST images in Figure~\ref{fig:hsttargets} can be seen, however spiral arm detail is only apparent for Harry and Neville. The black bars show $1~\rm{kpc}$ for scale along each dimension in each panel.} 
\label{fig:kcwitargets}
\end{figure*}

\section{Sample and observations}\label{sec:two}

\subsection{Sample Selection}\label{sec:sample}

We observed four disk-dominated galaxies with KCWI at the Keck Observatory, Hawai'i, USA, on the 13th December 2018. These were selected from a larger well-studied sample of $101$ disk-dominated galaxies with luminous, unobscured Type 1 AGN first identified in SSL17 ($\langle z \rangle = 0.129$). This parent sample was constructed from galaxies in the SDSS \citep{york00} Data Release 8 \citep{aihara11} imaging sample cross-matched with sources identified by \citet{edelson12} using multi-wavelength data from the Wide-field Infrared Survey Explorer \citep[WISE;][]{wright10}, Two Micron All-Sky Survey \citep[2MASS;][]{skrutskie06}, and ROSAT all-sky survey \citep[RASS;][]{voges99}. The disk-dominated morphologies were assigned by expert review of the SDSS imaging (see \citealt{simmons13} and SSL17), and were all later confirmed using images from an \emph{HST} snapshot survey with broadband imaging using ACS WFC (programme ID HST-GO-14606, PI: Simmons), which were reduced using the standard pipeline. \emph{HST} images showing the disk-dominated nature of our four targets, including spiral arms and bar features, along with the bright point source of the unobscured central AGN, are shown in Figure~\ref{fig:hsttargets}. Black hole masses for this sample were originally estimated by SSL17 using the relation between black hole mass and the FWHM and luminosity in the broadened $H\alpha$ emission line from \cite{greene05}. 

$58$ galaxies within this sample showed broadened blueshifted \oiii~components in their SDSS $3"$ fibre spectra. From this detection of a blueshifted component in the the spectra we know that there is \emph{some} outflowing material from the AGN within the $3"$ diameter central SDSS fibre, however this may not capture the full luminosity or extent of the outflow. The $12$ brightest galaxies in the blushifted \oiii~$5007\rm{\AA}$ spectral component were observed using narrowband filters on the Shane-3m telescope from 12-14th May 2018 at the Lick Observatory, California, USA. The results of this work are described in S19. We then selected $4$ of these targets to observe with KCWI; Harry, Padma, Neville and Theodore (continuing the naming convention used in S19; see Table~\ref{table:coords} for more details). These targets were visible from Mauna Kea in December 2018 and had an appropriate redshift to ensure \oiii~was in the wavelength range of KCWI. 

\subsection{KCWI observations}\label{sec:obs}
\rowcolors{1}{lightgray}{}
\begin{table}
\centering
\caption{Co-ordinates of the four disk-dominated AGN hosts observed with KCWI. }
\label{table:coords}
\begin{tabular}{lcccc}
\hline
Name  & SDSS name & RA & Dec & z \\
\hline
Harry & J0813+5422 & 123.350 & 54.377 & 0.043 \\
Padma & J1012+1017 & 153.161 & 10.289 & 0.070 \\
Neville & J1034+3938 & 158.661 & 39.641 & 0.043 \\
Theodore & J1314+4218 & 198.715 & 42.305 &  0.073 \\ 

\hline
\end{tabular}
\justify
\end{table}

We observed the $4$ disk-dominated AGN host galaxies listed in Table~\ref{table:coords} using KCWI at the Keck Observatory on Mauna Kea, Hawai'i, USA during dark time over half the night of the 13th December 2018. The weather was clear and the resultant seeing was $1.1''$. 

Our observational setup was determined by the combination of our need for a large field of view, high spectral resolution to resolve the emission lines of interest (\oiii~and \hbeta), and spectral bandpass coverage wide enough to allow for good continuum measurements for continuum subtraction. We used KCWI's blue camera with the `KBlue' filter. The field of view was $33''$ x $20''$, with a pixel scale $[0.30, 0.68] ''/\rm{pixel}$ using 2x2 binning. Using KCWI's large slicer allowed us to cover the full extent of all the galaxies in a single pointing. We used the BH3 grating, which allowed us to cover both \oiii~and \hbeta~with a spectral resolution of $R = 4500$, suitable for tracing the high-velocity line emission in these sources. The targets were bright enough that we were not significantly affected by the somewhat reduced throughput of the BH3 grating. 

Three targets (Harry, Padma \& Neville) were observed for $2,700$ seconds ($45$ minutes), with Theodore observed for $3,600$ seconds ($60$ minutes) to ensure a signal-to-noise ratio (SNR) of at least 10 for each target in the \oiii~emission. An inspection of the data cubes reveals that this SNR was exceeded for each target.

\section{Data Reduction \& Analysis}\label{sec:data}

\subsection{KCWI data reduction}\label{sec:datared}
Each KCWI raw data cube was reduced using the Keck Data Reduction Pipeline (KeckDRP) written in IDL\footnote{Note that a \emph{Python} Data Reduction Pipeline is now available for Keck data; see \url{https://kcwi-drp.readthedocs.io}}. The pipeline has 8 stages: a basic CCD reduction (bias and overscan subtraction, gain-correction, trimming and cosmic ray removal), dark subtraction, geometric transformation, flat-field correction, sky subtraction, data cube generation, atmospheric refraction correction and a flux calibration. The standard stars used for flux calibration were G191-B2B  and Feige 34. The total integrated flux across the data cubes for each of the four targets is shown in Figure~\ref{fig:kcwitargets}.

\begin{figure*}
\centering
\includegraphics[width=0.985\textwidth]{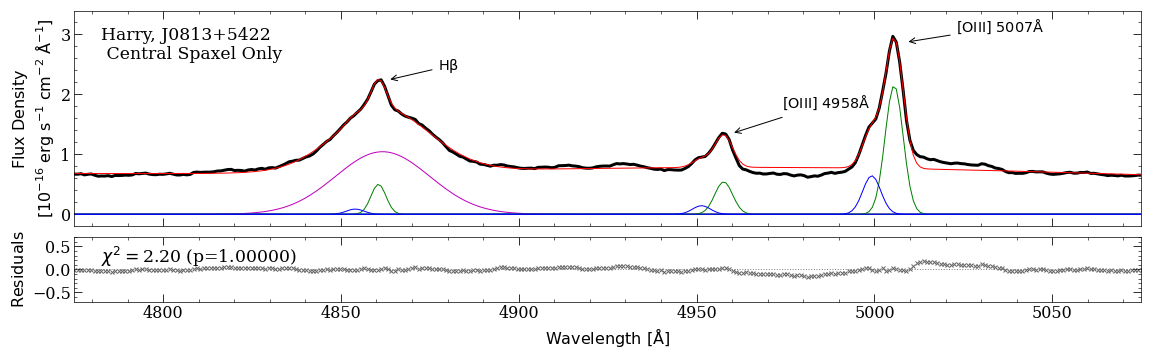}
\includegraphics[width=0.985\textwidth]{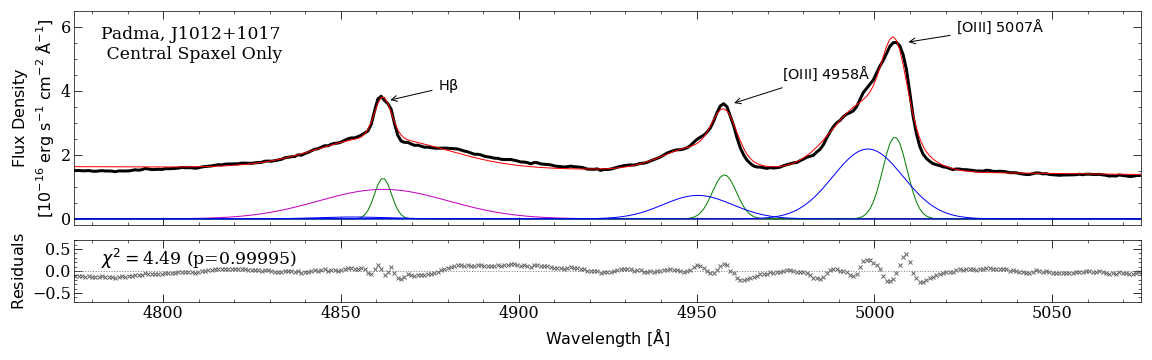}
\includegraphics[width=0.985\textwidth]{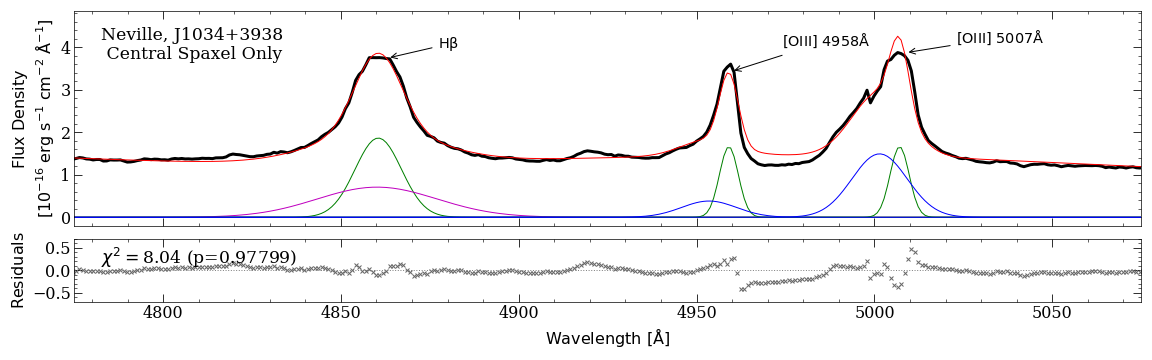}
\includegraphics[width=0.985\textwidth]{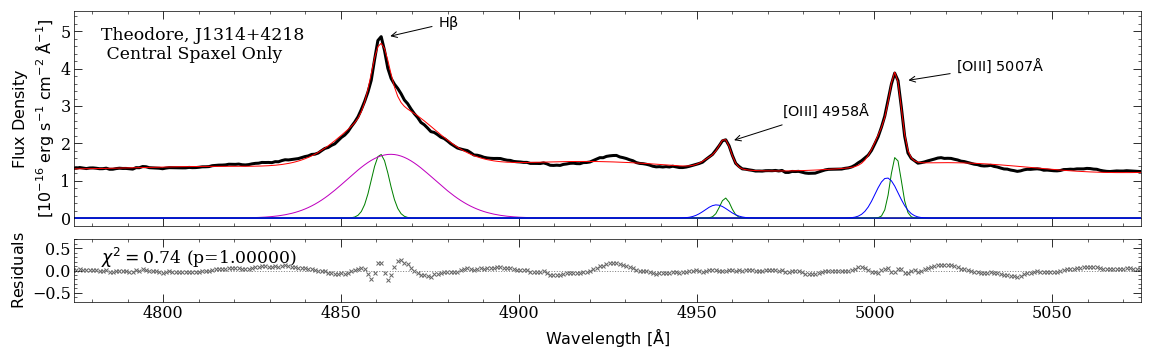}
\caption{The spectrum (black) and fit (red) to the brightest, central spaxel for each source. The individual components for each emission line are shown by the coloured lines (offset to 0). Each source was fitted with 2 components {\refchange(narrow in green and broad in blue)} for the \oiii~$4958\rm{\AA}$ and $5007\rm{\AA}$ emission lines, and with 3 components (narrow in blue, broad in green, and broad line region in magenta) for the \hbeta~emission line. Note that only Harry and Padma needed all three H$\beta$ components to fit to the brightest, central spaxel. The residual between the spectrum and the fit is shown below, with the $\chi^2$ value and corresponding p-value for a model with 21 degrees of freedom (amplitude, velocity and velocity dispersion for each of the 7 components). } 
\label{fig:specfits}
\end{figure*}

\subsection{Spectral fitting}\label{sec:specfit}

Once the reduced data cubes were obtained using the KeckDRP, we used the \emph{Python} module \texttt{ifscube}\footnote{\url{https://ifscube.readthedocs.io/}} to fit spectral features in the wavelength range probed by KCWI. {\refchange Systemic velocities were first determined using the peak of the \hbeta~emission in the central spaxel pre-decomposition\footnote{{\refchange Upon inspection of the final fits, the peak of the overall \hbeta~emission in the central spaxel coiincided with the peak of the narrow \hbeta~emission, see Figure~\ref{fig:specfits}}}, since stellar absorption lines were not available to us due to the Type 1 AGN nature of these systems (\citealt{RW18} show how \hbeta~is a good proxy for stellar absorption lines with an average velocity shift of $-9\pm^{41}_{45}~\rm{km}~\rm{s}^{-1}$).} Initially the flux, velocity and velocity dispersion of H$\beta$, \oiii~$4958\rm{\AA}$ and $5007\rm{\AA}$ were fitted with two components each, with one component required to have a broader velocity dispersion. After inspection of the spectra and the initial spectral fits, it was apparent that the central $H\beta$ emission was dominated by emission from the broad line region (BLR) of the AGN, and that the H$\beta$ and \oiii~narrow components were not kinematically linked, suggesting that the narrow \oiii~emission was ionised by the central AGN alone, rather than extended star formation in each source. 

We therefore reperformed the fits with three components for $H\beta$ (narrow, broad which was kinematically tied to the broad \oiii~components, and a BLR) and once again two components each for \oiii~$4958\rm{\AA}$ and $5007\rm{\AA}$(narrow and broad), with the narrow \oiii~components no longer kinematically tied to the narrow $H\beta$ component. The BLR component is also not kinematically tied to the narrow $H\beta$ component. The fits to the central spaxel for each source are shown in Figure~\ref{fig:specfits}, clearly showing the need for a BLR $H\beta$ component along with the obvious blueshifted outflows in \oiii. Only Harry and Padma (top panels of Figure~\ref{fig:specfits}) needed three components in H$\beta$ (narrow, BLR and outflow) in the central spaxel.

Note since these are Type 1 AGN we only expect to detect a blueshifted outflow component due to the effects of dust \citep{fischer13, muller11, baewoo14}. Indeed \cite{RW18} found that blueshifted \oiii~is more frequently detected than redshifted \oiii~by a factor of $3.6$ in Type 1 AGN (as opposed to a factor of 1.08 for Type 2 AGN) due to projection and orientation effects.

\begin{figure*}
\centering
\includegraphics[width=0.99\textwidth]{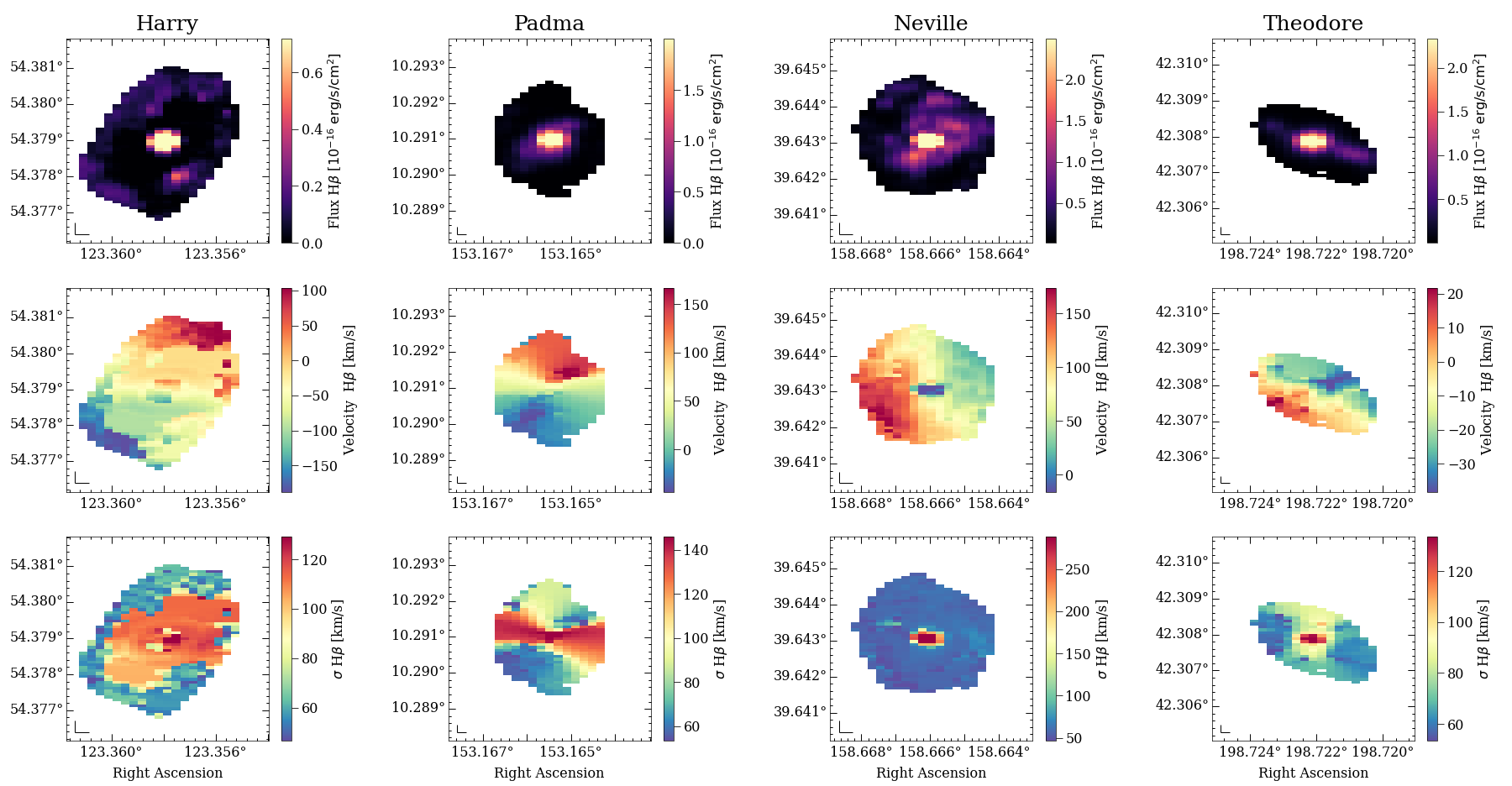}
\caption{The fit to the H$\beta$ narrow emission for the four targets observed with KCWI, showing the integrated flux (top; with an arcsinh stretch), velocity (middle; relative to the systemic velocity) and velocity dispersion, $\sigma$ (bottom). Pixels are masked if the flux is below 3 standard deviations. Note that the KCWI spectral resolution (and therefore the minimum resolvable $\sigma$ value) is $\sim60~\rm{km}~\rm{s}^{-1}$. The bars show $1~\rm{kpc}$ in each panel for scale.} 
\label{fig:hbetafour}
\end{figure*}

\begin{figure*}
\centering
\includegraphics[width=0.99\textwidth]{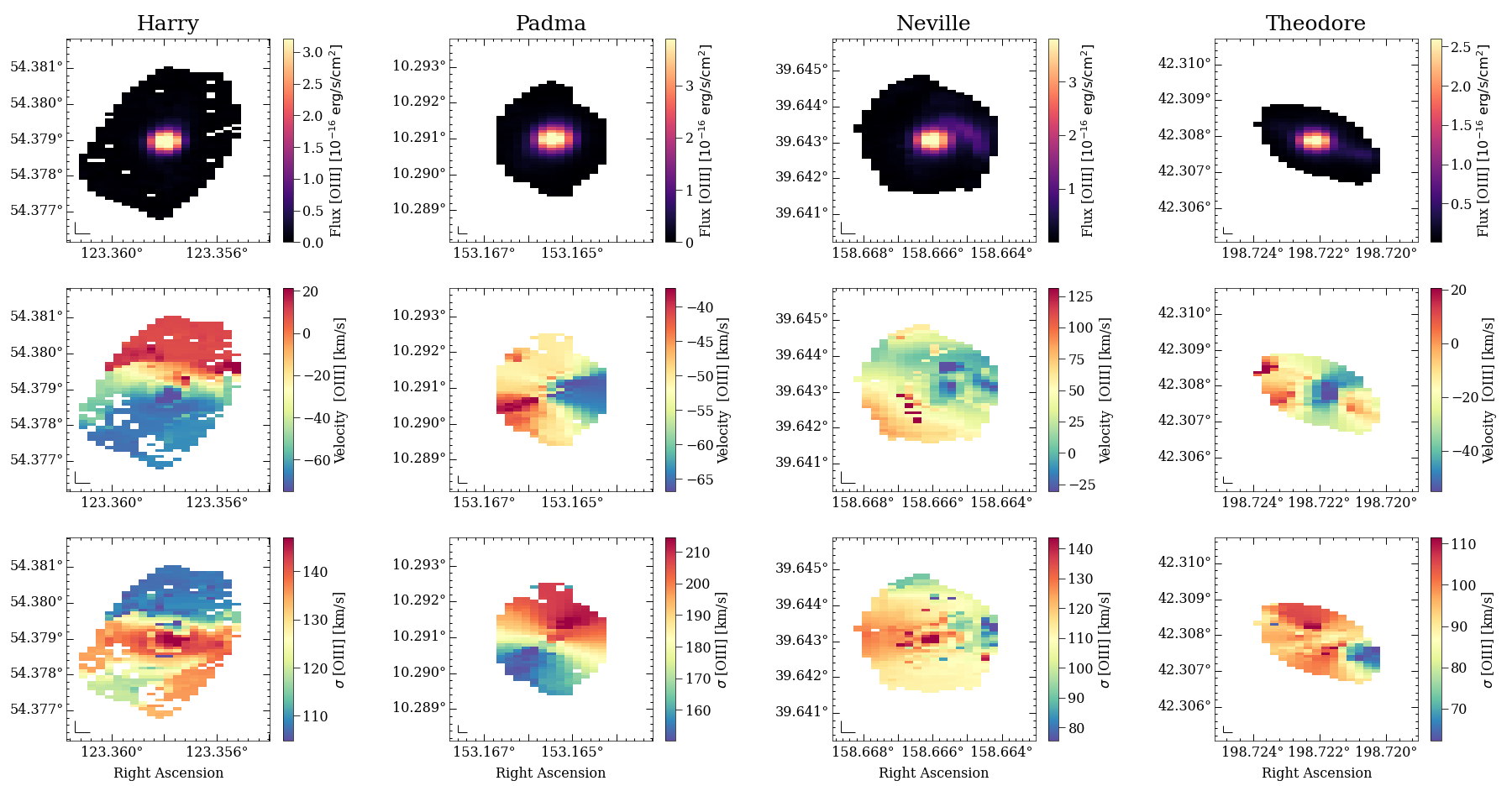}
\caption{The fit to the narrow \oiii~emission for the four targets observed with KCWI, showing the integrated flux (top; with an arcsinh stretch), velocity (middle; relative to the systemic velocity) and velocity dispersion, $\sigma$ (bottom). Pixels are masked if the flux is below 3 standard deviations. Note that the KCWI spectral resolution (and therefore the minimum resolvable $\sigma$ value) is $\sim60~\rm{km}~\rm{s}^{-1}$. The bars show $1~\rm{kpc}$ in each panel for scale.} 
\label{fig:oiiinfour}
\end{figure*}

\begin{figure*}
\centering
\includegraphics[width=0.99\textwidth]{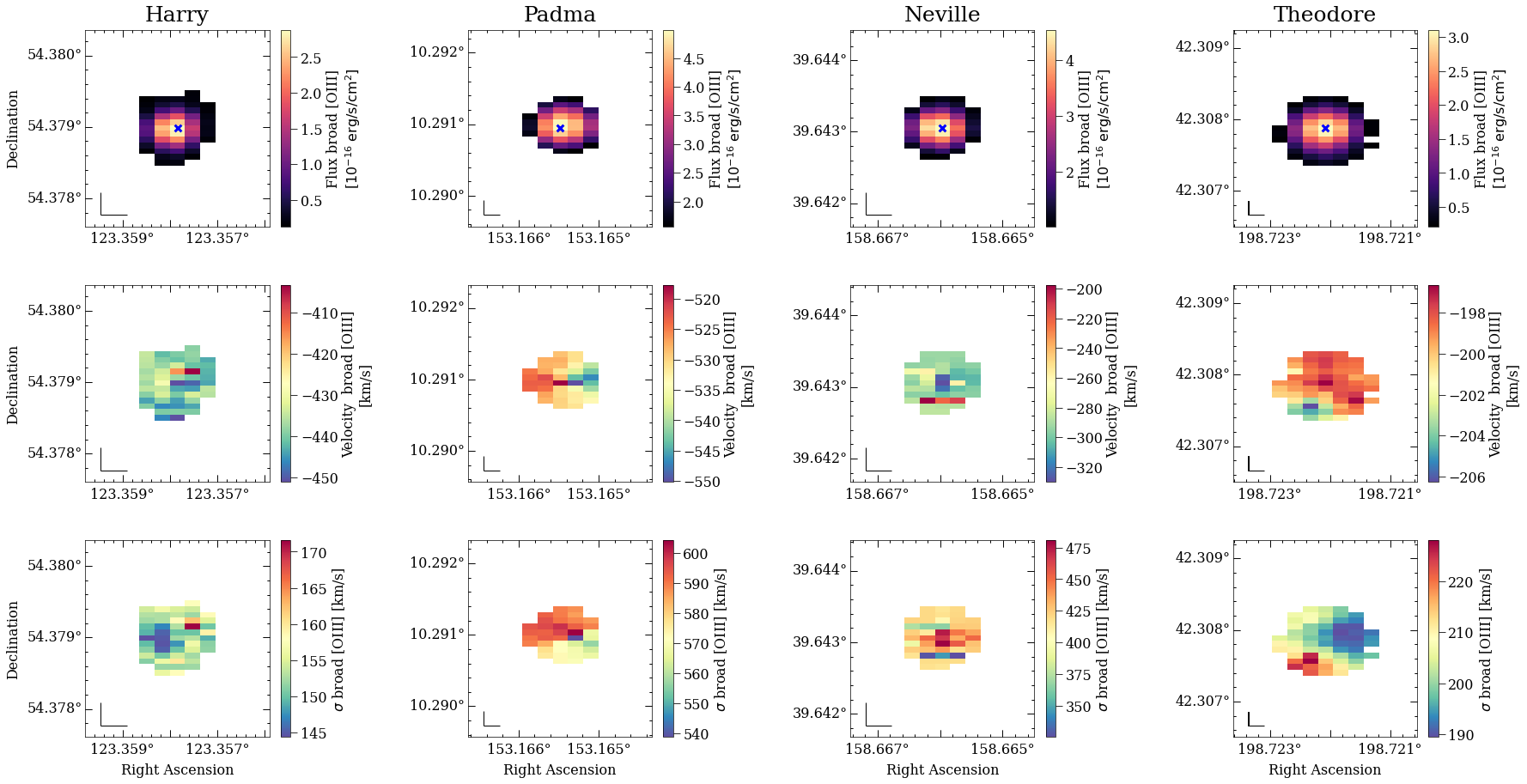}
\caption{The fit to the broadened \oiii~emission for the four targets observed with KCWI, showing the integrated flux (top; with an arcsinh stretch), velocity (middle; relative to the systemic velocity) and velocity dispersion, $\sigma$ (bottom). Note that the KCWI spectral resolution (and therefore the minimum resolvable $\sigma$ value) is $\sim60~\rm{km}~\rm{s}^{-1}$. The bars show $1~\rm{kpc}$ in each panel for scale; note the difference in scale to Figures~\ref{fig:hbetafour} \&~\ref{fig:oiiinfour}. Pixels are masked if the flux is below 3 standard deviations. In the top panels, the blue cross denotes the brightest point in the \oiii~narrow emission flux. For Padma, the position of the brightest outflow ionised emission is offset from the position of the brightest narrow emission ionised by the central AGN (marked by the blue cross). Note that the KCWI spatial resolution, combined with the ground based seeing, limits any further conclusions on the geometry or morphology of the outflows in these systems.} 
\label{fig:oiiiwfour}
\end{figure*}

The integrated flux, velocity and velocity dispersion of the narrow H$\beta$ emission are shown in Figure~\ref{fig:hbetafour} {\refchange with the top panels showing some of the structure in each system}. In Figure~\ref{fig:oiiinfour}, the integrated flux, velocity and velocity dispersion of the narrow \oiii~component is shown, assumed to be ionised by the central AGN (although note that Neville does show some extended narrow \oiii~emission presumably due to star formation along a spiral feature). Similarly, Figure \ref{fig:oiiiwfour} shows the integrated flux, velocity and velocity dispersion of the broad \oiii~components, assumed to be ionised by the AGN outflow. 

\begin{figure*}
\centering
\includegraphics[width=0.99\textwidth]{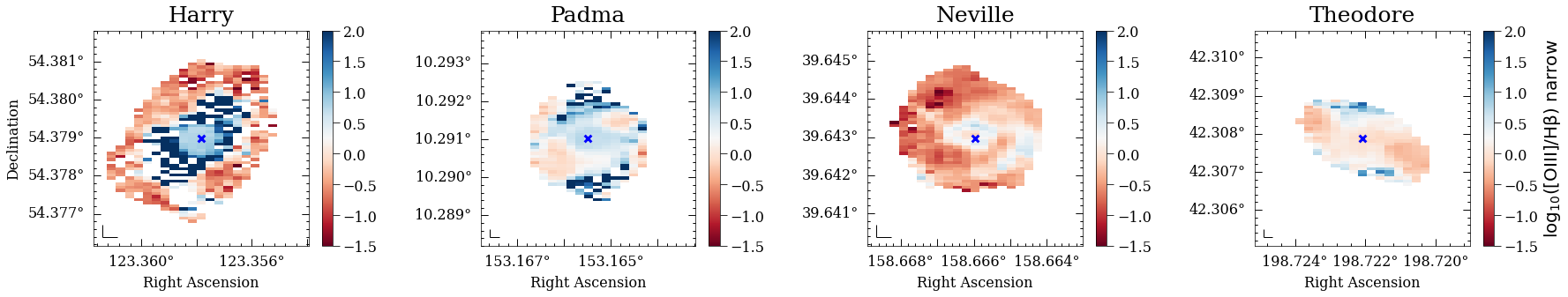}
\includegraphics[width=0.99\textwidth]{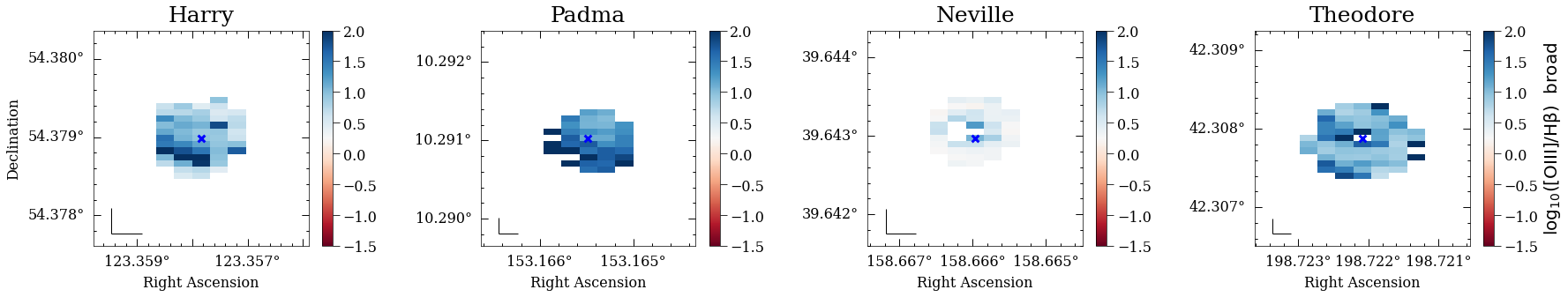}
\caption{The ratio of narrow (top) and broad (bottom) \oiii/H$\beta$ emission for each target. Note the change of scale between the two rows; the scale bars show $1\rm{kpc}$ in each panel. Here we use only the flux from the broad $H\beta$ component ionised by the outflow, and not from the $H\beta$ BLR component in these plots (note that the central spaxels for Neville and Theodore do not have outflow ionised H$\beta$ emission; see Figure~\ref{fig:specfits}). The colour bars are scaled between the typical ranges on a BPT diagram; star formation ionised emission typically has $\log_{10} [OIII]/H\beta \lesssim 0$ and AGN ionised emission typically has $\log_{10} [OIII]/H\beta \gtrsim 0$ {\refchange \citep{kewley01, kewley06}}. All of our sources have high broad \oiii/\hbeta~values, indicating that the outflows are ionized by the AGN.} 
\label{fig:ratios}
\end{figure*}


\subsection{Calculating \textsc{[OIII]} outflow rates}\label{sec:calcgasmass}

The fluxes shown in Figure~\ref{fig:oiiiwfour} enable a measurement of the outflow luminosity, $L$\oiii~(knowing the redshift of each target), which can then be used to calculate a gas mass in the outflow following the method outlined in \cite{carniani15}:
\begin{multline}\label{eq:carni}
M_{\rm{[OIII]}} = 0.8 \times 10^8~M_{\odot} ~\times \\ \left( \frac{C}{10^{[O/H] - [O/H]_{\odot}}} \right) \left( \frac{L[\rm{O}\textsc{iii}]}{10^{44}~\rm{erg}~\rm{s}^{-1}} \right) \left( \frac{n_e}{500~\rm{cm}^{-3}} \right)^{-1}
\end{multline}

where $n_e$ is the electron density, $[O/H] - [O/H]_{\odot}$ is the metallicity relative to solar, and $C = <n_e>^2 / <n_e^2>$. Here $<n_e>^2$ is the volume averaged electron density squared and $<n_e^2>$ is the volume averaged squared electron density. This method requires some simplifying assumptions regarding the nature of the outflowing gas, particularly the temperature, metallicity and density of the gas. The largest source of uncertainty when determining the mass outflow rate is the electron density, $n_e$. Typically, the \sii~emission is used to determine $n_e$ (although see \citealt{davies20} for a study showing that \sii~underestimates $n_e$), however the wavelength of \sii~is not probed by KCWI for these four targets.  

However, there is no general agreement on the best value of 
$n_e$ to use, with conflicting estimates across the literature for AGN at different redshifts. The long assumed value of $n_e = 100~\rm{cm}^{-3}$ has recently been challenged by \citet[][$700 < n_e < 3000~\rm{cm}^{-3}$]{perna17} and \citet[][$n_e \sim 10^5~\rm{cm}^{-3}$]{villar15}. Recent IFU studies have shown that $n_e$ can also vary spatially across a galaxy, for example \cite{mingozzi19} find a wide range of electron densities from $50-1000~\rm{cm}^{-3}$, with regions of high density concentrated in localized regions (which then dominate the total flux), while the rest of the regions in the galaxy have a much lower electron density.  In the outflows themselves, \cite{mingozzi19} find a median $n_e\sim250~\rm{cm}^{-3}$. This is an issue which plagues all such studies on AGN outflows since assuming a larger value of $n_e$ can lead to an underestimate of the gas mass present and vice versa.  We chose to use $n_e = 500~\rm{cm}^{-3}$ in order to be consistent with \cite{carniani15}. {\refchange However, we note that taking the extremes in $n_e$ found by \citet[][$50-1000~\rm{cm}^{-3}$]{mingozzi19} in comparison to the $n_e=500~\rm{cm}^{-3}$ value we use in this study, would result in outflow values either 10 times larger ($n_e =50~\rm{cm}^{-3}$) or two times smaller ($n_e=1000~\rm{cm}^{-3}$). In the absence of spatially resolved information of the electron densities for the 4 galaxies in this study, using an average value of $n_e=500~\rm{cm}^{-3}$ is therefore a reasonable choice.}

We also assume a gas solar metallicity, $[O/H] = [O/H]_{\odot}$. Since we are assuming a single value of $n_e$ and solar metallicity, the first term of Equation~\ref{eq:carni} reduces to unity. Note we do not include an uncertainty on $n_e$ when calculating an error on $M_{\rm{gas}}$ (or for the geometry of the system or volume filling factor), we propagate only the background noise and Poisson noise from the total flux (estimated using {\tt photutils.calc\_total\_error} function\footnote{\url{https://photutils.readthedocs.io/}}). 

We also investigate the kinematics of the outflow, including  the velocity of the outflow. Since the velocities and velocity dispersions measured for the broad \oiii~component (shown in Figure~\ref{fig:oiiiwfour}) only account for the velocity of the outflow along the line of sight, {\refchange whereas in reality the outflows will have a spread of observed radial velocities that will be lower than the actual bulk velocity of the outflow. The actual outflow velocity across 3-dimensions is best approximated by the most blueshifted velocity in the observed velocity distribution \citep{leung19}. A common parameter to measure this bulk velocity of the outflow is the maximum velocity , $v_{\rm{[OIII]}}$, determined as:
\begin{equation}\label{eq:velocity}
v_{\rm{[OIII]}} = |\Delta v_{\rm{max}}| + 2\sigma_{\rm{broad,[OIII],max}},
\end{equation}
where $|\Delta v_{\rm{max}}|$ is the maximum difference in the velocity of the narrow and broad \oiii~components, $\sigma_{\rm{broad,[OIII],max}}$ is the maximum velocity dispersion of the broad \oiii~component. The relation in Equation~\ref{eq:velocity} is defined by the properties of a normal distribution which is used to model the emission line velocity profiles \citep[see][]{rupke13}}. Not taking into account the line of sight effects on the velocity will result in an underestimate of the mass outflow rate (see Equation~\ref{eq:outflow}).

The physical extent of the outflow is also a key measurement for determining the scale over which these outflows will impact on the galaxy. We calculated the extent, $\rm{r}_{\rm{max}}$, as the most distant spatial extent of the broadened emission away from the central AGN (assumed to be the brightest pixel in the flux of the integrated \oiii~narrow emission shown in Figure~\ref{fig:oiiinfour}, with the location highlighted by the blue crosses in the top panels of Figure~\ref{fig:oiiiwfour}). We deconvolved our estimate of $\rm{r}_{\rm{max}}$ using an estimate of the seeing from observations of the standard star Feige 34. Not performing such a deconvolution results in an overestimate of the maximum physical extent and therefore an underestimate of the mass outflow rate (see Equation~\ref{eq:outflow}). 

Combining the velocity and physical extent allows for a calculation of the timesacle of the outflow:
\begin{equation}\label{eq:timescale}
t_{\rm{out}}~[\rm{yr}] = \bigg( \frac{\rm{r}_{\rm{max}}}{\rm{km}} \bigg)  \bigg( \frac{\rm{v}_{\rm{[OIII]}}}{\rm{km}~\rm{yr}^{-1}} \bigg)^{-1} .
\end{equation}
The mass outflow rate is then calculated in the following way:

\begin{equation}\label{eq:outflow}
\bigg(\frac{\dot{\rm{M}}_{\rm{out}}}{\rm{M}_{\odot}~\rm{yr}^{-1}} \bigg) = B \bigg( \frac{\rm{M}_{[OIII]}}{\rm{M}_{\odot}} \bigg) \bigg( \frac{\rm{t}_{\rm{out}}}{\rm{yr}} \bigg)^{-1}.
\end{equation}

Note that this method assumes that the outflow rate is constant over the time that the outflow has been active, $t_{\rm{out}}$. A factor of B between $1-3$ is typically applied to account for the geometry of the outflows \citep{harrison18}. For example, for a spherical outflow a factor of $B=3$ would be employed, whereas a biconical outflow covering only 1/3 of a sphere would need a factor of $B=1$. Given that our AGN host galaxies are disk-dominated and are assumed to be feeding the AGN through secular processes from the disk from the same angular momentum vector, we presume the outflow will not be spherical (see S19 and \citealt{npk12}) and therefore use a conservative value of $B=1$ throughout this work. This assumption may result in an underestimate of the outflow rate in these systems. 

\rowcolors{1}{lightgray}{}
\begin{table*}
\centering
\caption{Properties of the 4 disk-dominated AGN with outflow rates calculated from the extent and flux of \oiii~in spectral observations taken with KCWI. {\refchange We list black hole masses, $\log_{10}$ $[\rm{M}_{\rm{BH}}$/$\rm{M}_{\odot}]$, the \oiii~luminosity of the broad outflow component, $\log_{10}$ $[\rm{L}_{\rm{OIII}}$/$\rm{erg}~\rm{s}^{-1}]$, the Eddington ratio of the AGN, $\lambda_{\rm{Edd}}$, the accretion rate of the AGN, $\dot{m}$ (see Equation~\ref{eq:bhmdot}), the mass in the outflow,  $[\rm{M}_{\rm{OIII}}$/$\rm{M}_{\odot}]$ (see Equation~\ref{eq:carni}), the bulk outflow velocity,  $v_{\rm{max},[OIII]}$ (see Equation~\ref{eq:velocity}), the maximum radial extent of the outflow, $r_{\rm{max}}$ (see Section~\ref{sec:calcgasmass}), the outflow rate, $\dot{\rm{M}}_{\rm{out}}$ (see Equation~\ref{eq:outflow}), and the timescale of the outflow, $\rm{t}_{\rm{out}}$ (see Equation~\ref{eq:timescale}).}}
\label{table:rates}
\begin{tabular*}{\textwidth}{Cp{2.0cm}Cp{1.5cm}Cp{1.5cm}Cp{1.0cm}Cp{1.1cm}Cp{1.5cm}Cp{1.5cm}Cp{1.0cm}Cp{1.5cm}Cp{1.25cm}}
\hline
Name  & $\log_{10}$ $[\rm{M}_{\rm{BH}}$/$\rm{M}_{\odot}]*$ & $\log_{10}$ $[\rm{L}_{\rm{OIII}}$/$\rm{erg}~\rm{s}^{-1}]$ & $\lambda_{\rm{Edd}}$* & $\dot{m}$* $[\mathrm{M_{\odot}\,yr^{-1}}]$ & $\log_{10}$ $[\rm{M}_{\rm{OIII}}$/$\rm{M}_{\odot}]$ & $v_{\rm{max},[OIII]}$ $[\rm{km}~\rm{s}^{-1}]$ & $r_{\rm{max}}$ $[\rm{kpc}]$ & $\dot{\rm{M}}_{\rm{out}}$ $[\mathrm{M_{\odot}\,yr^{-1}}]\dagger$ & $\rm{t}_{\rm{out}}$ $\rm{[Myr]}$ \\
\hline
Harry & $6.56^{+0.13}_{-0.12}$ & $41.2\pm1.2$ & $0.08^{+0.33}_{-0.02}$ & $0.02^{+0.04}_{-0.01}$ & $5.1\pm0.1$ & $836\pm28$ & $0.6\pm0.3$ & $0.19\pm0.09$ & $0.6\pm0.3$\\
Padma & $7.62^{+0.14}_{-0.14}$ & $42.2\pm0.2$ & $0.20^{+0.45}_{-0.09}$ & $0.07^{+0.4}_{-0.3}$ & $6.03\pm0.09$ & $1710\pm6$ & $2.4\pm0.4$ & $0.7\pm0.1$ & $1.4\pm0.2$ \\
Neville & $6.30^{+0.12}_{-0.12}$ & $41.6\pm0.4$ & $0.86^{+2.90}_{-0.26}$ & $0.07^{+0.11}_{-0.04}$ & $5.5\pm0.1$ & $1316\pm29$ & $2.1\pm0.3$ & $0.18\pm0.03$ & $1.6\pm0.2$ \\
Theodore & $6.73^{+0.11}_{-0.11}$ & $41.6\pm0.6$ & $0.77^{+1.68}_{-0.35}$ & $0.06^{+0.04}_{-0.02}$ & $5.4\pm0.2$ & $675\pm18$ & $1.3\pm0.4$ & $0.12\pm0.04$ & $1.9\pm0.6$ \\ 

\hline
\end{tabular*}
\justify
\vspace{0.5em}
* Measurements from SSL17. Black hole masses are calculated using a virial assumption by measuring the full width half maximum of the broadened \ha ~component. SMBH accretion rates are calculated using bolometric luminosities inferred from WISE W3 magnitudes (see Section~\ref{sec:mdot}).\\
{\refchange $\dagger$ The quoted uncertainties on the outflow rates do not include an estimate of the uncertainty on the electron density, $n_e$ (see Section~\ref{sec:calcgasmass}). In this study we use a value of $n_e=500~\rm{cm}^{-3}$ to calculate the mass in the outflow to be consistent with \cite{carniani15}, but we note that taking the extremes in $n_e$ found by \citet[][$50-1000~\rm{cm}^{-3}$]{mingozzi19}, results in outflow rates either 10 times larger ($n_e =50~\rm{cm}^{-3}$) or two times smaller ($n_e=1000~\rm{cm}^{-3}$) than quoted here. The mean outflow rate of the four targets would therefore be in the range of $\langle\dot{M}_{\rm{out}}\rangle = 0.15-3~\rm{M_{\odot}}~\rm{yr}^{-1}$.}
\end{table*}

The kinetic energy outflow rate and momentum flux of the outflow can then be calculated as:
\begin{equation}\label{eq:kinout}
\dot{E}_{\rm{out}} = \frac{1}{2} \dot{M}_{\rm{out}} v_{\rm{[OIII]}}^2
\end{equation}
and
\begin{equation}\label{eq:momout}
\dot{P}_{\rm{out}} = \dot{M}_{\rm{out}}v_{\rm{[OIII]}}
\end{equation}

respectively.

\subsection{Black hole accretion rates}\label{sec:mdot}

The SMBH accretion rate can be inferred from the bolometric luminosity of the AGN, $L_{\rm{bol}}$;
\begin{equation}\label{eq:bhmdot} 
\dot{m} = L_{\rm{bol}}/\eta c^2,
\end{equation} 
where the radiative efficiency, $\eta =0.15$ (see \citealt{elvis02}). Bolometric luminosities were originally inferred by SSL17 for these four targets using the WISE W3 band magnitudes at $12\mu m$, by applying a correction from \cite{richards06}. It is possible that the W3 flux densities could be contaminated by star formation, however \cite{richards06} concluded that since there were minimal differences between their composite SEDs of Type 1 AGN around $\sim12\mu m$ this suggested minimal host galaxy contamination. Unlike for \oiii~which could still have some star formation contamination in the narrow component for our four targets (e.g. see top panel of Figure~\ref{fig:oiiinfour} for Neville). In addition, the normalisation factor used to convert $L_{\rm{[OIII]}}$ to $L_{\rm{bol}}$ is highly uncertain. While \cite{heckman04} suggest a normalisation factor of $\sim3500$, there is some debate in the literature over the correct value, with some arguing it is \oiii~luminosity dependent \citep[e.g.][estimate it ranges from 87-454]{lamastra09}. We therefore decided to use the bolometric luminosities previously calculated by SSL17 using the less problematic W3 flux densities.


\section{Results}\label{sec:results}

The top panels of Figure~\ref{fig:oiiiwfour} show the integrated flux in the broad \oiii~component which are used to calculate the gas masses, velocities, physical extents and outflow rates given in Table~\ref{table:rates}. The mean \oiii~gas mass in the outflow for the four targets is $\langle\rm{M}_{\rm{[OIII]}}\rangle = 5.5\pm0.2~\rm{M}_{\odot}$ (with a range of $5.1-6.03 ~\rm{M}_{\odot}$), with a corresponding mean outflow rate of $\langle\dot{M}_{\rm{out}}\rangle = 0.3\pm0.1~\rm{M}_{\odot}~\rm{yr}^{-1}$ (range $0.12-0.7~\rm{M}_{\odot}~\rm{yr}^{-1}$)\footnote{{\refchange Note that the uncertainties on these values do not include the uncertainties on the electron density $n_e$ (see Section~\ref{sec:calcgasmass}). In this study we use a value of $n_e=500~\rm{cm}^{-3}$ to be consistent with \cite{carniani15} in order to calculate the mass in the outflow, but we note that taking the extremes in $n_e$ found by \citet[][$50-1000~\rm{cm}^{-3}$]{mingozzi19}, results in outflow rates either 10 times larger ($n_e =50~\rm{cm}^{-3}$) or two times smaller ($n_e=1000~\rm{cm}^{-3}$) than quoted. The mean outflow rate of the four targets would therefore be in the range of $\langle\dot{M}_{\rm{out}}\rangle = 0.15-3~\rm{M_{\odot}}~\rm{yr}^{-1}$.}}. The outflows are substantial with a mean maximum radial extent of $\langle\rm{r}_{\rm{max}}\rangle = 1.6\pm0.4~\rm{kpc}$ (range $0.6-2.4~\rm{kpc}$), which is $\sim25\%$ of the galaxy Petrosian radius on average. {\refchange These extents are similar to those found in other AGN outflow studies, for example \citet{bae17} found that the mean outflow radius in their sample (20 Type 2 AGN at $z<0.1$) was $\sim1.8~\rm{kpc}$, \citet{harrison14} found a range in \oiii~outflow extents of $1.5-4.3~\rm{kpc}$ (16 Type 2 AGN $z<0.2$), and \cite{kang18} measured outflows ranging from $0.60-7.45~\rm{kpc}$ in size (23 Type 2 AGN $z<0.2$)}. Figure~\ref{fig:ratios} shows the resolved narrow and broad \oiii/H$\beta$ ratios and reveals how the outflows are ionized by the AGN in all four targets.

The gas mass values are consistent with those found by S19 using a narrowband imaging technique with the Shane-3m at the Lick Observatory, although are on average $\sim1$ dex larger. This is unsurprising given that S19 struggled to cleanly separate the broad and narrow emission using narrowband data (either due to extended star formation or subtraction of the central AGN PSF), and were only able to derive a lower limit on the gas mass for Neville. This suggests that the PSF subtraction dominated the uncertainty in the measurements of S19, resulting in an underestimate of the \oiii~gas masses. Note that the values initially quoted by S19 were affected by a standard star flux calibration error, and were on average overestimated by $2.6$ dex. This has since been corrected with an erratum (Smethurst et al. 2021; erratum submitted). We are not able to directly compare the velocities or maximum extents of the outflows (and therefore the outflow rates) derived as S19 used $|\Delta v_{\rm{max}}|$ rather than $v_{\rm{[OIII]}}$ and did not deconvolve their measurement of $r_{\rm{max}}$ (see Section~\ref{sec:calcgasmass}), both of which lead to an underestimate of the outflow rates. Note that $v_{\rm{[OIII]}}$, as used in this study is a more accurate representation of the maximum outflow velocity (see Section~\ref{sec:calcgasmass} and Equation~\ref{eq:velocity}).

Despite the many limitations to narrowband imaging, it does allow for a higher spatial resolution in order to discern the basic morphology and features of each outflow. The KCWI data in this study has low spatial resolution and does not allow us to draw any conclusions about the features of each outflow (the biggest limitation is the seeing, estimated at $1.1''$). The top panels of Figure~\ref{fig:oiiiwfour} reveal how the brightest pixel in the broadened \oiii~emission ionised by the outflow also coincides with the brightest pixel in the narrow \oiii~emission (ionised by the central AGN and star formation) for 3 of our sources (Padma has an offset). If there is structure to the outflows, it is lost due to the combination of the large pixel size of KCWI and the seeing. Therefore, in order to make any statements about the morphology of these outflows, more observations will be required with a higher spatial resolution IFU with AO capabilities (e.g. such as MUSE on the VLT).

\subsection{Harry (J0813+5422)}\label{subsec:harry}

Harry has the strongest bar feature of each of the four galaxies targetted in this study (as seen in Figures~\ref{fig:hsttargets} \& \ref{fig:kcwitargets}). The spiral features are picked up in the \hbeta~emission seen in the top left panel of Figure~\ref{fig:hbetafour}, with the velocity map revealing the ordered rotation in this feature. The narrow \oiii~emission, shown in Figure~\ref{fig:oiiinfour}, is centrally concentrated and shows some ordered rotation, suggesting this emission is ionised by a combination of the AGN and central star formation. The blueshifted, broadened \oiii~emission shown in Figure~\ref{fig:oiiiwfour} however, does not show clear rotation in the velocity map. Figure~\ref{fig:ratios} reveals how the central region and the outflow  have high \oiii/\hbeta~ratios, suggesting that the outflow is indeed ionised by the AGN. Table~\ref{table:rates} reveals that Harry has the lowest ionised gas mass, the lowest SMBH accretion rate and the lowest spatial extent of the four targets. This suggests Harry's outflow is relatively new, therefore it is unsurprising that Harry has the shortest timescale over which it is estimated to have been active of all four targets: $0.6~\rm{Myr}$ (see Table~\ref{table:rates}).

\subsection{Padma (J1012+1017)}\label{subsec:padma}

Figure~\ref{fig:hsttargets} reveals that Padma has a bar lens feature \cite[an oval-like structure along the bar major axis, see][]{athan15} surrounded by spiral structure. This spiral structure is not detected in the \hbeta~emission flux shown in Figure~\ref{fig:hbetafour}, however the corresponding \hbeta~velocity map shows the most ordered rotation of the four targets studied (and similarly for the narrow \oiii~velocity map in Figure~\ref{fig:oiiinfour}). The brightest point in the broadened \oiii~flux is offset from the brightest point in the narrow \oiii~flux (shown by the blue c ross in Figure~\ref{fig:oiiiwfour}). Padma has the largest ionised gas mass of all four targets, at an order of magnitude larger than Harry. Padma also has the largest SMBH mass, SMBH accretion rate, outflow velocity and physical extent ($2.4~\rm{kpc}$), leading to the largest outflow rate of the four targets of $0.7\pm0.1~\rm{M}_{\odot}~\rm{yr}^{-1}$. The ratio between the SMBH accretion rate and the outflow rate is therefore much larger, meaning more of the inflowing material is ejected in the outflow than is accreted by the SMBH (i.e. a higher mass loading factor, see \citealt{qui21} for example). 

\subsection{Neville (J1034+3938)}\label{subsec:neville}

Neville has prominent flocculent spiral features and a possible weak bar, as revealed by the HST imaging in Figure~\ref{fig:hsttargets}. Emission from this flocculent structure is identifiable in the \hbeta~emission flux shown in Figure~\ref{fig:hbetafour}, where clear rotational structure can be seen in the velocity map. The centre of Neville's \hbeta~velocity and velocity dispersion map show broadened emission with little rotation, suggesting the central \hbeta~emission is ionised by the AGN and not star formation, which is confirmed by the relatively high \oiii/\hbeta~ratios seen in Figure~\ref{fig:ratios}. Extended narrow \oiii~emission across a spiral feature can be seen in Figure~\ref{fig:oiiinfour}, suggesting ionisation by star formation is also present along with ionisation from the central AGN. S19 also reported extended \oiii~emission from Neville in their narrowband imaging data, resulting in an uncertain isolation of the emission ionised by the outflow alone. With one of the largest SMBH accretion rates, the SMBH is accreting at a similar order of magnitude to the measured outflow rate. The outflow has one of the highest velocities and physical extents ($2.1~\rm{kpc}$) after Padma.

\subsection{Theodore (J1314+4218)}\label{subsec:theodore}

Theodore has a strong bar feature with faint, loosely wound spiral arms emerging from the ends (as seen in HST imaging in Figure~\ref{fig:hsttargets}). Figure~\ref{fig:kcwitargets} reveals how only the bar feature is picked up in the KCWI observations. This is particularly apparent in the flux in the \hbeta~emission shown in Figure~\ref{fig:hbetafour}, which also reveals some rotational structure in the corresponding velocity map. This bar feature is also just noticeable in the narrow \oiii~emission (Figure~\ref{fig:oiiinfour}), suggesting ionisation due to ongoing star formation in the bar. This could also extend into the central regions of the galaxy as the narrow \oiii/\hbeta~ratio in the top right panel of Figure~\ref{fig:ratios} is low, suggesting ionisation dominated by star formation. However, the broad \oiii/\hbeta~ratios in the bottom panels of the same figure are high, suggesting the outflow is ionised by the AGN and not stellar winds. Like Neville, the SMBH accretion rate of Theodore is of the same order of magnitude as the outflow rate (a factor of just $\sim2$ difference). The resulting outflow has the lowest velocity of the four targets observed.

\section{Discussion}\label{sec:discussion}

Given that the targets we have observed in this study are all disk-dominated with little to no bulge component (see Figure~\ref{fig:hsttargets}), we assume the galaxy and the SMBH have co-evolved via a non-merger process \citep{walker96,hopkins12, martig12, tonini16, stevens16}. We must therefore consider which processes are able to drive an inflow of gas of at least $0.21-0.77~\rm{M}_{\odot}\rm{yr}^{-1}$ (to power both the accretion of the SMBH and the outflow) for an extended period of $0.6-1.9~\rm{Myr}$ (the time over which the outflows in our four targets have been active, see Table~\ref{table:rates} and Equation~\ref{eq:timescale}). 

Bars and spiral arms are long-lived morphological features and could therefore feasibly drive an inflow to the central regions of a galaxy over many $\rm{Gyr}$ \citep{fanali15, hunt18, jung18}\footnote{Note that these simulations only considered galactic scale inflows and did not consider how gas was transferred from kpc to sub-pc scales in the central regions. Therefore, these simulations don't provide estimates for the amount of gas that makes it to the AGN accretion disk itself, merely that which is transferred to the central gas reservoir.}. All four of our targets show clear spiral features (see Figure~\ref{fig:hsttargets}), with Harry and Theodore showing a strong bar feature, Neville a weak bar feature \citep{nair10b} and Padma a barlens feature \cite[an oval-like strucutre along the bar major axis, see][]{athan15}. Simulations suggest both bars and spiral arms can drive inflows at rates an order of magnitude larger than needed to power the combined outflow and SMBH accretion rates for all four targets \cite[$0.1-\rm{few}$ $M_{\odot}~\rm{yr}^{-1}$;][]{regan04, davies09, lin13,fanali15,slater19}. This order of magnitude difference is promising, since our simplifying assumption here is that the inflow must be at least enough to power both the SMBH accretion and the outflow, this means that the inflow would be sufficient to also fuel central star formation or contribute to the central gas reservoir \citep{tacconi86, boker03, bigiel08, leroy13, moreno21}. This suggests that bars and spiral arms would be capable of driving inflows which could sustain both the SMBH growth and an outflow from the AGN, while still contributing gas to the central gas reservoir of the galaxies.

S19 compared their AGN outflow rates and SMBH accretion rates to the results of \cite{bae17}, who studied a sample of $20$ nearby ($0.024 < z < 0.098$) Type 2 AGN with mixed morphologies (two of their sample are ongoing mergers) using the Magellan/IMACS-IFU and VLT/VIMOS-IFU\footnote{These IFUs had a large enough wavelength range to allow \cite{bae17} to empirically determine the column densities of the ionised gas, $n_e$, using the \sii~line ratio, unlike in this study with KCWI. They found a range of $54 < n_e < 854~\rm{cm}^{-3}$, with an average $n_e\sim360\pm230~\rm{cm}^{-3}$ which is similar to the value of $n_e=500~\rm{cm}^{-3}$ used in this study. B17 also used the $M -\sigma_*$ relation of \cite{park12} to derive black hole masses (rather than the virial assumption of \cite{greene05} as implemented by SSL17). In addition they calculated bolometric luminosities from the luminosity of the central narrow \oiii~emission \cite[see][]{heckman04}, as opposed to deriving them using the WISE W3 band at $12\mu m$ as implemented by SSL17. The reader is urged to bear these caveats in mind while the two studies are compared.}. Although we only have four targets in this study we can still make some comparisons to the \cite{bae17} sample. The velocities of the outflows in our sample are comparable to the \cite{bae17} sample (when calculated in the same way as Equation~\ref{eq:velocity}), with our four targets having higher velocities by a factor of $\sim1.35$ on average. However, the average outflow rates for our four targets are much lower than those of the merger powered \cite{bae17} sample, $\sim15$ times lower on average. However, the black hole accretion rates are larger in our four targets than the \cite{bae17} sample by a factor of $\sim3$ on average. This is in agreement with the findings of S19, who discussed the possibility that this scenario could be explained by higher spin of the SMBHs in the disk-dominated sample, following the hypothesis of \cite{npk12}.

Given that the outflow rates of the merger-grown \cite{bae17} sample are $\sim15$ times larger than the outflow rates of the four disk-dominated galaxies studied in this work, this suggests that the inflow rates funnelled by merger processes must be much larger than in secular processes. However, given the comparable accretion rates of the black holes powering the AGN, these inflows do not contribute to the growth of the black hole, but instead are used to power a large outflow which can have considerable impact on the surrounding galaxy. This supports the conclusions of \cite{mcalpine20}, who found using the EAGLE simulations, that mergers do not induce a significant amount of SMBH growth, instead finding that the majority of mass is accreted by the SMBH outside the merger period. Similarly \cite{martin18} showed using the Horizon-AGN simulation that only $\sim35\%$ of all of the matter contained in SMBHs by $z\sim0$ is a result of mergers (either major or minor). Combining these results with our findings here suggests that secular processes are responsible for the majority of SMBH growth, whereas mergers are responsible for the majority of outflowing material and the subsequent feedback on the galaxy. 

\begin{figure}
\centering
\includegraphics[width=0.475\textwidth]{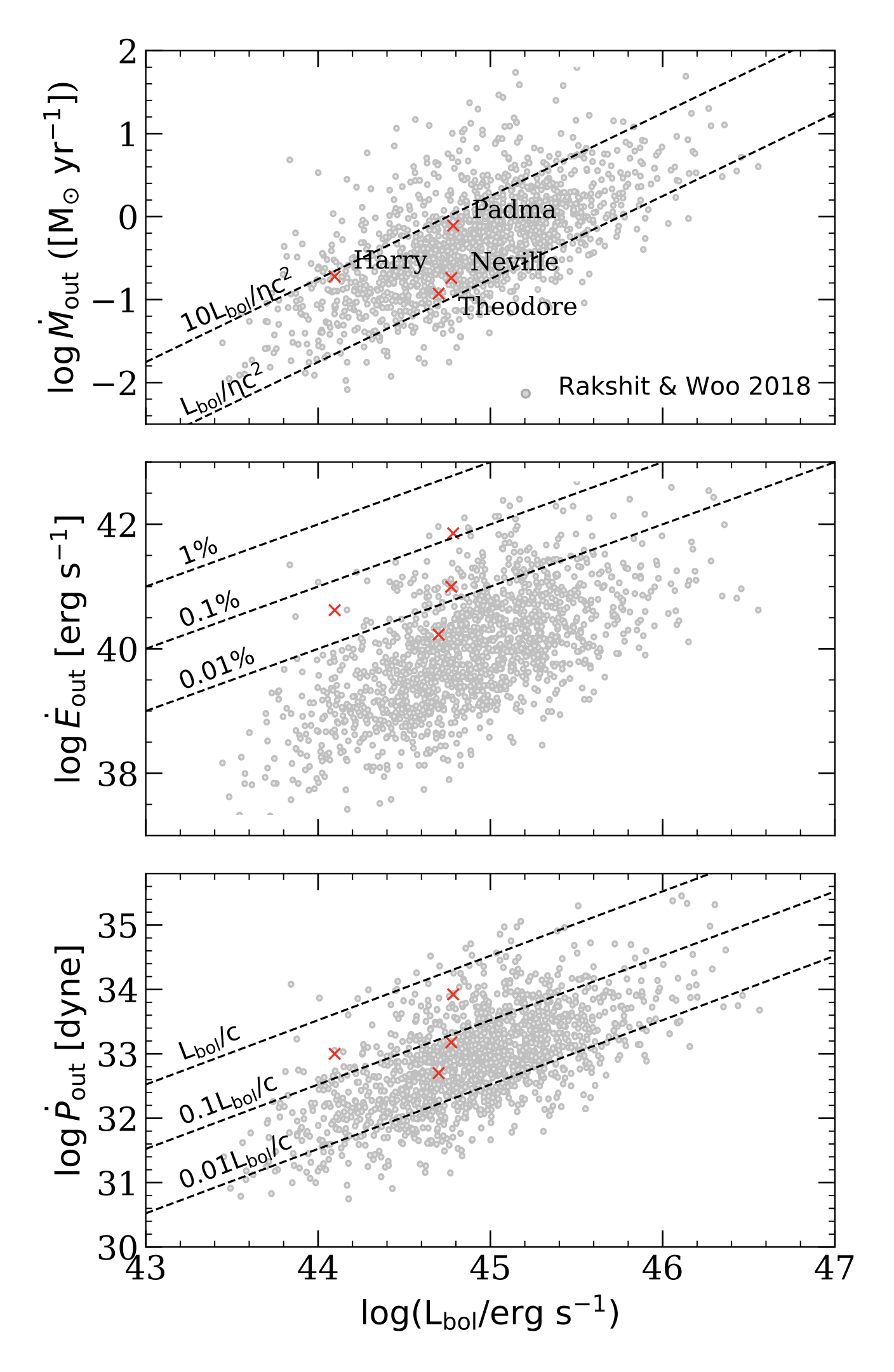}
\caption{The mass outflow rate (top), energy injection rate (middle), and momentum flux (bottom) against the AGN bolometric luminosity ($L_{\rm{bol}} = 3500~L_{[OIII]}$) for our four sources (red crosses). This figure is a recreation of Figure 11 from \protect\citet{RW18}; we compare our sources with their estimates for 5221 Type 1 AGNs from SDSS ($z<0.3$; shown by the grey circles). This figure shows how our secularly powered outflows are typical of low-redshift Type 1 AGN and that they have momentum conserving outflows.} 
\label{fig:energetics}
\end{figure}

We also compare the outflow rates, kinetic energy outflow rate and momentum flux of the outflow calculated for our sample to a sample of $\sim5000$ Type 1 AGN identified in SDSS from \cite{RW18}\footnote{Note that \cite{RW18} used SDSS spectra to determine outflow gas masses, which may miss some outflow flux outside the fibre (leading to a possible underestimate of the outflow rate) and inferred the physical extent of the outflow using an empirical relation with \oiii~luminosity from \cite{kang18}. {\refchange In addition, \cite{RW18} estimated bulk outflow velocities as $v_{out} = (\sigma_{\rm{broad,[OIII],max}}^2 + |\Delta v_{\rm{max}}|^{2})^{0.5}$, which is different from how we estimated the bulk velocities in this study (see Equation~\ref{eq:velocity}). Calculating our outflow velocities in this way results in lower values than quoted in Table~\ref{table:rates}, by $541~\rm{km}~\rm{s}^{-1}$ on average. This particularly affects the comparison of $\dot{E}_{out}$ which has a $v_{\rm{out}}^2$ dependency, leading to an average difference in $\log_{10}\dot{E}_{\rm{out}}$ of $\sim0.7$ dex (and $\sim0.34$ dex in $\log_{10}\dot{P}_{\rm{out}}$). Readers should bear these caveats in mind while comparing the results of this study with those from \cite{RW18} in Figure~\ref{fig:energetics}, however we note that these differences due to the alternate bulk outflow velocity estimate used do not account for the differences between our four targets and the Type 1 AGN population seen in Figure~\ref{fig:energetics}.}} in Figure~\ref{fig:energetics}. We find that the outflow rates of our four targets are comparable to the larger AGN population given their bolometric luminosities. However, given their larger velocities, this results in higher kinetic energy injection rates and momentum flux compared to the larger AGN population, but still within the typical range. This figure demonstrates that the secularly powered outflows of our four targets are typical of low-redshift Type 1 AGN. It is worth noting here that many AGN are found in non-merger systems (for example see \citealt{smethurst16, aird19}), with a wide-range of morphologies, which may also be fuelled by these same secular processes. Given that we find that our outflows and accretion rates are typical of the larger low-redshift AGN population, and given the results of simulations such as \cite{martin18} and \cite{mcalpine20}, it is possible that the majority of low-redshift AGN (both growth and outflows) are powered by secular processes. 

The momentum flux of the outflows allows us to probe whether the outflows are momentum-conserving \cite[i.e. winds driven by radiation pressure][]{thompson15, costa18} or energy-conserving \cite[i.e. driven by fast, small-scale winds][]{faucher12, costa14}. The average ratio of $\log_{10}[c\dot{P}_{\rm{out}}/L_{\rm{bol}}] = -0.91\pm0.08$ suggests that these outflows are momentum-conserving. If the ratio was higher than unity, then an extra boost of momentum from an energy-conserving wind (which does work on the surrounding material, therefore increasing the momentum of the large-scale wind) would be required. The measurements of the kinetic energy injection rate allow us to probe the physical driver of the outflows observed in our four targets. For example the ratio of $\dot{E}_{\rm{out}}/L_{\rm{bol}}$ is between $0.004\%-0.12\%$ for our targets, meaning that the AGN is energetically sufficient to drive the observed outflows. This is in agreement with the high \oiii/\hbeta~ratios seen in Figure~\ref{fig:ratios} suggesting that the outflows are ionised by the AGN rather than star formation. Such low values of $\dot{E}_{\rm{out}}/L_{\rm{bol}}$ are often interpreted as outflows which are incapable of impacting their surrounding galaxy through AGN feedback. Many theoretical works claim that only those outflows with $\dot{E}_{\rm{out}}/L_{\rm{bol}} \gtrsim 0.5-5\%$ are capable of quenching galaxies \citep{dimatteo05, hopkins10, harrison18}; however Figure~\ref{fig:energetics} shows how the majority of low-redshift AGN do not achieve such high efficiencies, with the majority $<1\%$. 

To determine whether the outflows of our four targets will have an effect on their host galaxies, we first compare the velocity of each outflow to the escape velocity of the galaxy at a radius equal to the maximum extent of the outflow. We assume an $n=1$ Sersic profile to model the light distribution in each galaxy and calculate the fraction within the most distant spatial extent of the outflow, $r_{max}$. We then assume a constant mass-to-light ratio in order to work out the total stellar mass of the galaxy within that radius, $M_{*,r<r_{\rm{max}}}$. The escape velocity of the galaxy at the maximum extent of each outflow is then calculated as $v_{esc, gal} = (GM_{*,r<r_{max}}/r_{max})^{0.5}$, assuming spherical symmetry. The average $v_{[OIII]}$ for the four targets in our sample is $1134\pm 205~\rm{km/s}$, which is $\sim30.5$ times larger than the average escape velocity of the galaxy. We can therefore assume that these outflows, despite their relatively lower rates, will escape the galactic potential and cause AGN feedback to the galaxy by driving gas out of the central regions, or cause feedback to the galactic halo through heating the intergalactic medium (note the large radial extent of the outflows in these four targets of $0.6-2.4~\rm{kpc}$, which is $\sim25\%$ of the galaxy Petrosian radius on average).

In order to determine whether the outflows are impacting each galaxy, we would need an estimate of the resolved SFR (e.g. from H$\alpha$ and/or D$_n4000$). The wavelength range of KCWI does not cover these spectral features in the redshift range of these sources; an IFU with a larger wavelength range would be necessary to quantify the feedback efficacy. Since these are Type 1 AGN the SFRs derived from SDSS spectra are also unreliable due to contamination from the AGN. However, it is worth noting that these four targets have galaxy $u-r$ colours\footnote{Calculated in a `donut' shaped aperture by removing the SDSS PSF magnitude from the Petrosian magntiude.} in the range $1.7-2.5$ ($\pm0.1$; although note this is not the case for the parent sample of disk-dominated galaxies, see Section~\ref{sec:sample}) and would therefore be classified as either Green Valley or Red Sequence galaxies \citep{baldry04, smethurst16}. 

In addition, SSL17 demonstrated how these disk-dominated systems lay on the typical galaxy stellar mass-SMBH mass correlation (i.e. within the scatter), suggesting that non-merger co-evolution of galaxies with their SMBH is possible. Therefore, if \emph{both} merger-driven and non-merger-driven SMBH growth lead to co-evolution, this suggests that this co-evolution is regulated by feedback in both scenarios. Confirming whether AGN outflows in disk-dominated galaxies are powerful enough to cause feedback is therefore of great importance for our understanding of galaxy evolution through co-evolution. An IFU with a larger wavelength range (to cover e.g. $\rm{H}\alpha$ in order to probe the SFR), high spatial resolution (to more accurately resolve the regions impacted by the outlow) and better seeing (this is the biggest limiting factor using KCWI) would allow for a more detailed study on the feedback effects of outflows powered by secular processes in these disk-dominated systems. For example, an IFU such as MUSE on the Very Large Telescope (VLT), used with adapative optics, would be ideal for this science case.

\section{Conclusion}\label{sec:conc}

We have observed four disk-dominated galaxies hosting luminous AGN with KCWI, an IFU available at the Keck observatory. These galaxies are assumed to have their evolution (and therefore their SMBH growth) dominated by non-merger processes due to their lack of central bulge (see Figure~\ref{fig:hsttargets}). 

We performed spectral fits to each of the reduced data cubes from KCWI and detected blueshifted broadened \oiii~components in all four targets with \oiii/\hbeta ratios indicative of ionisation by the AGN. With these spectra we were able to spectrally isolate the broadened \oiii~emission from the narrow \oiii~emission ionised by the central AGN (see Figures~\ref{fig:oiiinfour} \&~\ref{fig:oiiiwfour}). From these fits we calculated the integrated flux in \oiii~$4958\rm{\AA}~\&~5007\rm{\AA}$ across each target and from this calculated the total ionised gas mass in the outflow (see Equation~\ref{eq:carni}). From the maximum extent of the outflow (see top panels of Figure~\ref{fig:oiiiwfour}) and the bulk velocity of the outflow we were able to estimate the outflow rate (see Equation~\ref{eq:outflow}), energy injection rate and momentum flux for these four systems. Our conclusions are as follows:
\begin{enumerate}
    \item The outflow rates of the four targets range from $0.12-0.7~\rm{M}_{\odot}~\rm{yr}^{-1}$, with corresponding SMBH accretion rates in the range $0.02-0.7~\rm{M}_{\odot}~\rm{yr}^{-1}$. The velocities, outflow rates, kinetic energy injection rate and momentum flux of these secularly powered outflows are all typical of other low-redshift AGN outflows in the literature.
    \item Secular processes such as funnelling of gas by bars and spiral arms are more than capable of providing enough gas to power both the accretion and outflow rates measured in this study, with simulations suggesting they can power inflows an order of magnitude larger than the combined SMBH accretion and AGN outflow rates observed. This suggests that a significant amount of inflow funnelled to the centre by secular processes, will not necessarily be used for SMBH growth or AGN outflows, but will contribute to the central gas reservoir of the galaxy.
    \item The maximum radial extent of the outflows is substantial, ranging from $0.6-2.4~\rm{kpc}$, which is on average $\sim25\%$ of the galaxy Petrosian radius. 
    \item The outflow velocities in all of our AGN exceed ($\sim30$ times larger on average) the escape velocity of the galaxy at the maximum radial extent of the outflow. This suggests that these outflows will have a feedback effect on their galaxies, perhaps expelling gas from the central regions or heating the surrounding halo. This suggests that if the co-evolution of SMBHs and galaxies is possible through both merger and non-merger driven growth, then AGN feedback may be responsible for regulating this co-evolution in both scenarios. Further spectral observations using an IFU with a larger wavelength range and higher spatial resolution will be needed to quantify the resolved feedback efficacy of these outflows.
    \item We find that the outflow rates in the merger-powered AGN sample of \cite{bae17} are $\sim51$ times larger than in our four disk dominated targets, whereas the SMBH accretion rates are $\sim3$ times lower. This is in agreement with the findings of \cite{smethurst19b} who attributed this to the hypothesised spin up of SMBHs due to a secular feeding mechanism. 
\end{enumerate}

Combining our results with the conclusions of recent simulations \citep[e.g.][]{martin18, mcalpine20} suggests that secular processes are responsible for the majority of SMBH growth over cosmic time. A higher spatial resolution IFU study, supported by adaptive optics, of the larger parent sample of these four disk-dominated galaxies would allow for a more detailed study on the SMBH growth processes and AGN feedback effects of outflows powered by secular processes in these disk-dominated systems.

\section*{Acknowledgements}

RJS gratefully acknowledges funding from Christ Church, Oxford. BDS gratefully acknowledges current support from a UK Research and Innovation (UKRI) Future Leaders Fellowship (MR/T044136/1) and past support at the time of KCWI proposal and observing from the National Aeronautics and Space Administration (NASA) through Einstein Postdoctoral Fellowship Award Number PF5-160143 issued by the Chandra X-ray Observatory Center, which is operated by the Smithsonian Astrophysical Observatory for and on behalf of NASA under contract NAS8-03060.

This research is based on observations made with the NASA/ESA Hubble Space Telescope obtained from the Space Telescope Science Institute, which is operated by the Association of Universities for Research in Astronomy, Inc., under NASA contract NAS 5–26555. These observations are associated with program HST-GO-14606. 

This research made use of Astropy,\footnote{http://www.astropy.org} a community-developed core Python package for Astronomy \citep{astropy13, astropy18} and the affiliated {\tt ccdproc} package \citep{ccdproc}.

The data presented herein were obtained at the W. M. Keck Observatory, which is operated as a scientific partnership among the California Institute of Technology, the University of California and the National Aeronautics and Space Administration. The Observatory was made possible by the generous financial support of the W. M. Keck Foundation.

The authors wish to recognize and acknowledge the very significant cultural role and reverence that the summit of Maunakea has always had within the indigenous Hawai'ian community.  We are most fortunate to have the opportunity to conduct observations from this mountain.

Funding for the Sloan Digital Sky Survey IV has been provided by the Alfred P. Sloan Foundation, the U.S. Department of Energy Office of Science, and the Participating Institutions. SDSS acknowledges support and resources from the Center for High-Performance Computing at the University of Utah. The SDSS web site is \url{www.sdss.org}.

SDSS is managed by the Astrophysical Research Consortium for the Participating Institutions of the SDSS Collaboration including the Brazilian Participation Group, the Carnegie Institution for Science, Carnegie Mellon University, the Chilean Participation Group, the French Participation Group, Harvard-Smithsonian Center for Astrophysics, Instituto de Astrof\'isica de Canarias, The Johns Hopkins University, Kavli Institute for the Physics and Mathematics of the Universe (IPMU) / University of Tokyo, Lawrence Berkeley National Laboratory, Leibniz Institut für Astrophysik Potsdam (AIP), Max-Planck-Institut f\"ur Astronomie (MPIA Heidelberg), Max-Planck-Institut für Astrophysik (MPA Garching), Max-Planck-Institut f\"ur Extraterrestrische Physik (MPE), National Astronomical Observatories of China, New Mexico State University, New York University, University of Notre Dame, Observat\'orio Nacional / MCTI, The Ohio State University, Pennsylvania State University, Shanghai Astronomical Observatory, United Kingdom Participation Group, Universidad Nacional Aut\'onoma de M\'exico, University of Arizona, University of Colorado Boulder, University of Oxford, University of Portsmouth, University of Utah, University of Virginia, University of Washington, University of Wisconsin, Vanderbilt University and Yale University.

\section*{Data Availability}

The data underlying this article will be shared on reasonable request to the corresponding author.

\bibliographystyle{mn2e}
\bibliography{refs}  

\bsp	
\label{lastpage}
\end{document}